# The Feng Shui of Visualization: Design the Path to SUCCESS and GOOD FORTUNE

韩畅 (Chang Han)*
University of Utah

Andrew Mcnutt†
University of Utah

**Abstract**

Superstition and religious belief system have historically shaped human behavior, offering powerful psychological motivations and persuasive frameworks to guide actions. Inspired by Feng Shui—an ancient Chinese superstition—this paper proposes a pseudo-theoretical framework that integrates superstition-like heuristics into visualization design. Rather than seeking empirical truth, this framework leverages culturally resonant (superstitious) narratives and symbolic metaphors as persuasive tools to encourage desirable design practices, such as clarity, accessibility, and audience-centered thinking. We articulate a set of visualization designs into a Feng Shui compass, reframing empirical design principles and guidelines within an engaging mythology. We present how visualization design principles can be intepreted in Feng Shui narratives, discussing the potential of these metaphorical principles in reducing designer anxiety, fostering community norms, and enhancing the memorability and internalization of visualization design guidelines. Finally, we discuss Feng Shui visualization theory as a set of cognitive shortcuts that can exert persuasive power through playful, belief-like activities.

## 1 Superstition, Religion, Feng Shui, and Visualization

> "...being frequently driven into straits where rules are useless, and being often kept fluctuating pitiably between hope and fear... they (human) are consequently, for the most part, very prone to credulity."
>
> — Benedict de Spinoza, *A Theologico-Political Treatise, Preface Part 1.*

Superstitions, in all their curious forms, are widely used to (mentally) combat uncertainty and hardship [24]. The tendency to embrace superstition are found to be relevant with multiple factors, including the level of education, political and cultural background, and occupation (it is especially common in professionals with high uncertainty, such as sports), among others [23]. Superstition is so widespread that it is even observed across different species—one famous experiment, *"Superstition in the pigeon"*, shows that a caged, starved pigeon will superstitiously repeat certain actions, even though the release of the food hopper was actually controlled by humans and had no relation to its behavior [21].

Superstition can take on many different forms; among them, Feng Shui is a highly developed one that is deeply embedded in Chinese history [25]. Although it has been labeled as a superstition since the modern era due to a lack of scientific proof, Feng Shui remains quite popular in many regions around the world to this day [25]. This paper does not take the position that Feng Shui is true or correct, nor does it adopt an oppositional stance toward Feng Shui theory. It does not take a strictly neutral position either, but rather deliberately avoids this question. Instead, it focuses on the power of Feng Shui theory, as a culturally embedded tool, to exert real world influence.

*e-mail: changhan@sci.utah.edu
†e-mail: andrew.mcnutt@utah.edu

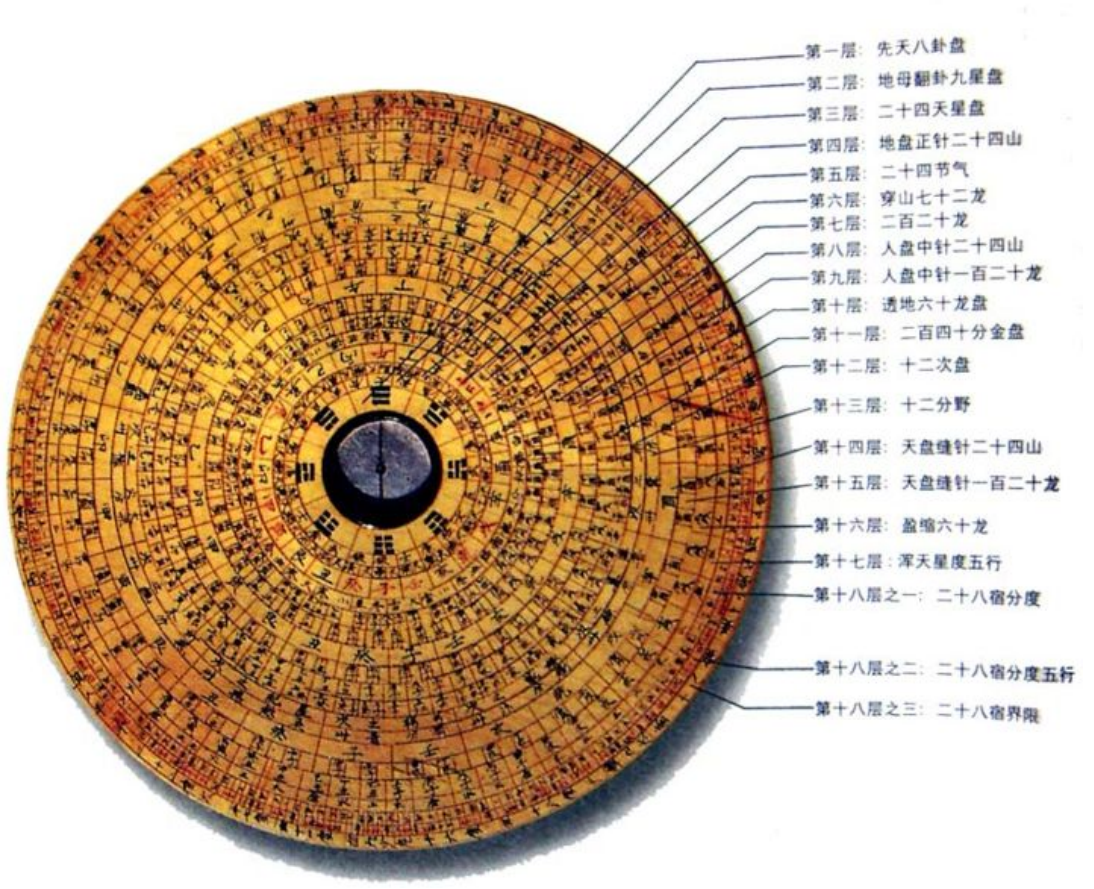

Figure 1: Luo Pan (罗盘), a magnetic compass also known as the Feng Shui compass. It contains directional and temporal information. Some rings are rotatable. Credit: Pcae18 (Wikipedia)

Such influence can operate on many dimensions (considered good or bad depending on the value system). Practicing superstition is not merely grasping at straws when desperate; it has been found to yield psychological benefits with positive outcomes, helping people improve performance in golf, motor dexterity, memory, and even anagram games [7]. In fact, the form and development of superstition may arise directly from materialistic benefits. David Hume, in his *The Natural History of Religion*, links superstition as a source of polytheism, which he regarded as the primary and most primitive religion of humanity. Arguably, if we are to consider religion as a form of well-developed superstition, Marvin Harris's cultural materialism view might provide an interesting explanation for how this development occurred—Harris explained certain religious food taboos as materially realistic choices shaped by economic and environmental realities [9]. This paper similarly adopts a cultural materialist perspective: some, though certainly not all, superstitious traditions have roots in lived realities. This is at least partially true for Feng Shui, where one study found that its application in southeast China resulted in reasonable site selections that reduced the effects of natural disasters such as floods [15].

There exists some similar explorations in broader computer science field, people have made use of western notions of occult as means to interrogate our understanding of algorimtic practices. Browne and Swift [2] make use of ritual centered on a Ouija board to defamiliarize viewers with neural networks. Similarly, Börütecene [1] draw on Ouija as means to enrich the design process. McNutt et al. [17] employ tarot to drive a visualization recomendation system as a means to question the kinds of faith that we put into recommender systems. Byrne et al. [4] colllect these and instances as examples of spooky technology, which they use as framework to reflect on relationships with technology.

To the best of our knowledge, there is no previous work that blends religious or superstitious beliefs into visualization. However, there are investigations into how visualization is used in religious contexts to depict the human body—called *data cultures* [5]. Visualization—depending on how you define it—has long played a role in religious settings. Feng Shui itself is full of visual elements and symbolic rituals that could arguably be considered a form of visualization. This paper, in a reversal of current approaches, uses a religious/superstitious theory to reflect on the design of visualization. To borrow the trendy jargon, this is a move from VIS4Religion to Religion4VIS. Specifically, this paper attempts to reinterpret empirically studied visualization design guidelines [11, 13, 22, 8] through the lens of Feng Shui-style beliefs.

**Why do we need a superstition theory of Feng Shui in visualization research?** There is no doubt that visualization is a scientific domain, so we are not arguing that we truly "need" it. Rather, the argument is that there is no harm in having it—just like the way many people approach Feng Shui with a mix of skepticism and belief: "It can not hurt to believe—just in case. (Pascal's wager)" Taking it a step further, the position of this paper is that it would be interesting to see how such a theory might evolve in people's minds, and whether it could also wield persuasive power or provide psychological benefits. Lastly, we honestly do not know if visualization as a research community will survive into the next century—but Feng Shui as a practice probably will. (Of course, this is just gibberish—it is hardly fair to judge the value of something by how long it sticks around in human history.) Also, perhaps the passionate stance on maximizing the "data-ink ratio" or the disdain for pie charts isn't just about design. Maybe they gesture toward an implicit belief system, like Feng Shui, that guides certain visualization practices through implicit normative principles.

**Positionality:** Both authors are not Feng Shui experts. The first author grew up in a cultural context where Feng Shui was present and is familiar with its basic concepts. The second author has no connection to Feng Shui.

## 2 The Wind and Water: Feng Shui Basics

**Disclaimer:**

> There exists a range of debates, opinions, and doctrines of thought regarding Feng Shui [16]. The authors of this paper do not claim to be Feng Shui experts. In fact, the discussion of Feng Shui in this work is primarily based on the perspectives presented by Xixian Yu (a late Professor in Peking University who researched Feng Shui) in his public lecture series [26], which is intended as an introductory course rather than advanced content.

Feng Shui (风水, literally "wind and water") is an ancient Chinese superstition that has relevance to daoism [25]. Common perceptions are that it is about interior design or furniture, but Feng Shui is actually a compound theory that operates on multiple levels, serving both as a guide for action and a lens through which reality is interpreted. On a grand scale, it has been used to explain the rise and fall of dynasties and the strategic placement of capital cities. On a smaller scale, it dictates the layout of homes and tombs, the arrangement of furniture, and even the proportions of doors and windows [26]. Feng Shui is also called Kan Yu (堪舆, literally "examining the landscape") [16].

Feng Shui views the environment as a holistic system in which harmony among space, time, and human factors is essential to achieving an auspicious (ji, 吉) state and avoiding inauspiciousness (xiong, 凶). Central to this practice is the theory of the Five Elements—metal, wood, water, fire, and earth—and their dynamic cycles of generation (sheng, 生) and overcoming (ke, 克), which describe how elements support or restrain one another. Auspiciousness is assessed by examining the interactions among spatial orientation, temporal rhythms, and personal attributes. Spatial factors include the layout and directionality of structures, while temporal aspects involve the energetic qualities of specific times, marked by the Heavenly Stems and Earthly Branches (a time system). The human dimension is anchored in a person's birth data, or Eight Characters (ba zi, 八字), which reflect their elemental constitution. Feng Shui aims to align these three dimensions—space, time, and self—by identifying and resolving imbalances, especially clashes (fan chong, 犯冲) that disrupt the flow of energy (qi) and lead to misfortune. Through spatial adjustments, favorable timing, and strategic choices of materials or colors, practitioners aim to regain balance and enhance well-being.

Over time, Feng Shui theory underwent significant development and diversified into different schools of thought, including modern efforts to reinterpret it through a scientific lens [16]. The sheer volume and complexity of knowledge that has accumulated around Feng Shui goes beyond the scope of this paper. To set boundaries, the discussion of Feng Shui here is based primarily on the content of a public course offered by Peking University, taught by the late Professor Xixian Yu [26], an authority on Feng Shui research in China. We also refer to concepts found in an ancient text traditionally attributed to Guo Pu of the Jin Dynasty, written approximately 1,500 years ago. His work, The Book of Burial (葬书), introduced key ideas such as qi (energy), the field of qi, and the principle that all things should strive toward harmonious balance. These are the foundational Feng Shui principles adopted in this paper.

## 3 Luo Pan of Visualization Design

The Luo Pan (Figure 1), or Feng Shui compass, is one of the most important tools in Feng Shui practice. Each ring (Pan) on the compass has a name and is used to measure a particular aspect of Feng Shui, often relating to time (year, month, or day), location, or the attributes of artifacts (such as the five basic elements). Feng Shui masters use the compass to measure and calculate the Feng Shui of an environment[16]. Different Feng Shui doctrines have different types of Luo Pan, reflecting their own theories and worldviews.

Drawing inspiration from the Luo Pan, we present the Luo Pan of visualization design to help people measure the Feng Shui of their visualizations. It features seven rings—far fewer than a typical Feng Shui compass, which may have 21 rings. We consider this a new doctrine of Feng Shui, just getting started, and expect that future development may further enrich this compass. Each ring is given a traditional Feng Shui name, with assigned new meanings to them, grounded in empirically tested visualization design guidelines.

We present an overview of the Visualization Luo Pan; we will dive into the details and discuss some use cases in the following sections. Note that the design rationale discussed here is presented in terms of visualization design theories and their correlation to Feng Shui, but the actual usage of the compass does not involve any "scientific" language—instead, it uses Feng Shui narratives. The compass begins with the Pan (ring) of the people (人盘). In Feng Shui, people are one of the most important considerations, as the calculation of harmony is closely tied to human attributes. In the Visualization Luo Pan, the Pan of the people covers everything about those involved in the visualization—primarily the target audience, but also the designer itself. To achieve harmony, people must come first. What is their experience level? Are there accessibility concerns? Do they have particular preferences? Next comes the Pan of the earth (地盘). In Feng Shui Visualization, this represents the foundation of the visualization: the dataset. This involves considering the type, scale, cleanliness, and other attributes of the dataset to achieve harmony. Following that is the Pan of the sky (天盘), which stands for the driving force of visualization: the task—ranging from low-level cognitive tasks like making comparisons to high-level goals like pleasing executives. Afterwards is the Pan of Na Jia (纳甲), which relates to the construction of visualization charts—the visual encoding channels. This is followed by the Pan of the five elements (五行), representing the color and style of the chart. Finally, there is the outer Pan (外盘), the physical foundation of any visualization: the display medium or device. To reach harmony, each Pan must be carefully considered and aligned—otherwise, misfortune (in the form of poor design) may soon follow.

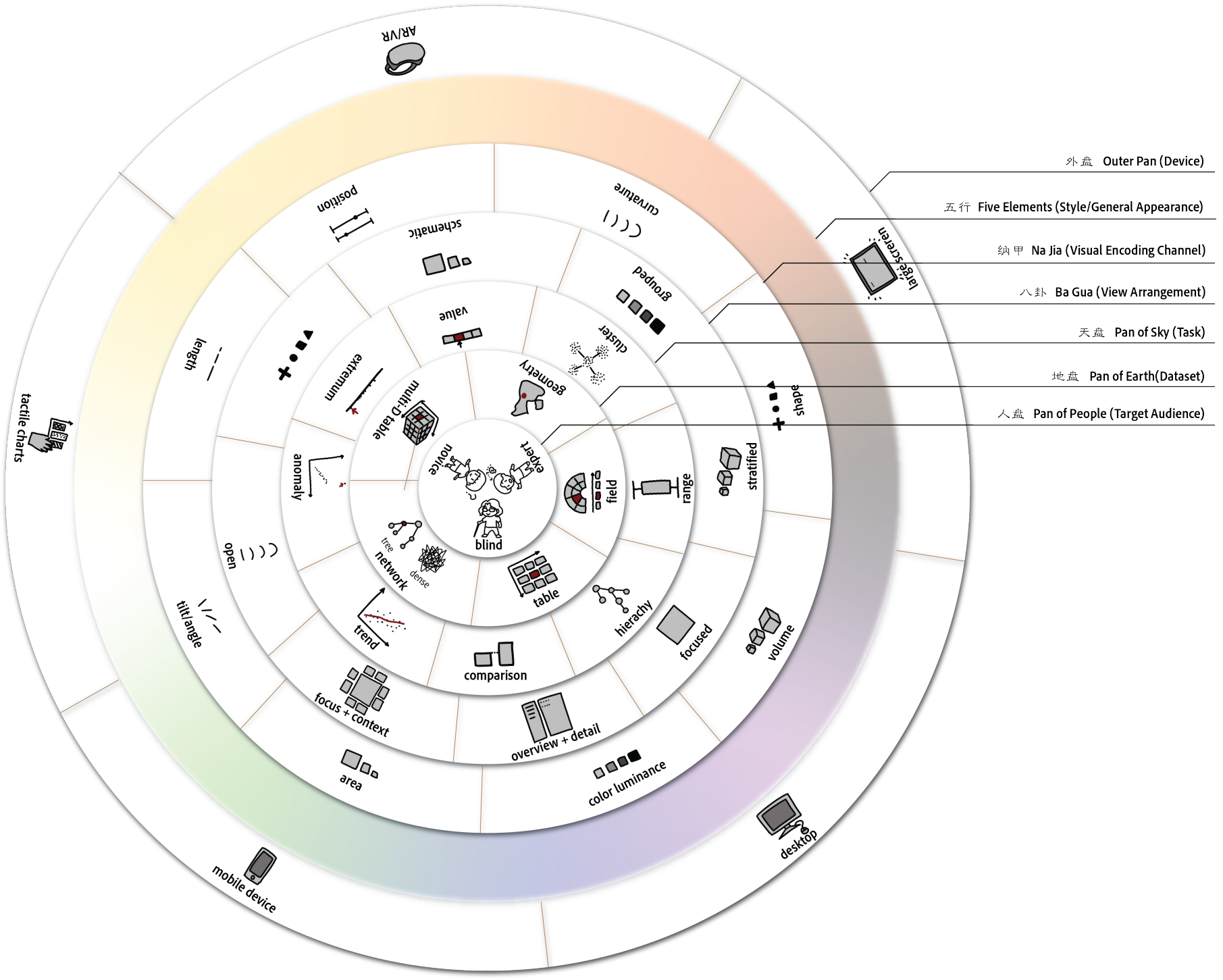

Figure 2: The Feng Shui Luo Pan (Compass). Each ring is assigned both a name in Feng Shui theory and a design aspect in terms of visualization. It should be noted that the sectors shown on each ring are merely illustrative and do not imply that these are the only divisions present. In fact, the content represented by each ring is broadly defined and can be adapted as needed. A high resolution version is availble in the appendix. A live rotatble version is available on https://hconhisway.github.io/FengShuiPan/.

## 4 Visualization and Feng Shui: Both are Human-Centered Endeavors

One major goal of visualization is to leverage the human visual system so that data can be perceived "intuitively." But intuitive for whom? The target audience, of course. Effective visualization designs should be human-centered. There is a growing body of work focusing on questions such as who the audience is [3], how to evaluate visualization literacy [14, 19, 6], and how to better support different groups in using visualizations [18].

Feng Shui cares about people, too. In Feng Shui one of the most important concept is clash or conflict (fan chong, 犯冲), which arises when attributes of elements—such as directions, time markers (Heavenly Stems (天干) and Earthly Branches (地支)) are in opposition to personal attributes. Such clashes are believed to disrupt the flow of qi, leading to inauspiciousness, such as illness, misfortune, or interpersonal tensionn. For example, certain directions may clash with an individual's birth sign, or the time of an event might be deemed incompatible with the elemental composition of the space or the people involved. Therefore, a key function of Feng Shui is to identify and mitigate these imbalances or conflicts. By adjusting spatial arrangements, choosing favorable dates, or even selecting appropriate colors and materials, practitioners aim to restore harmony among the interacting forces and thereby improve the well-being and fortune of the people involved.

In short, according to Feng Shui theory, the quality of an environment depends on who lives in it. Calculating Feng Shui thus often involves considering the attributes of the person involved. That's why we put the Pan of People at the center of the Luo Pan. To achieve harmony, the final visualization design must be suitable for the people involved in it, otherwise then will fan chong (clash). For example, an overly complex or information-dense chart will fan chong with audience who are novices. Flashy visual effects will fan chong with domain scientists who use the visualization for data analysis—conversely, charts in a concert poster will fan chong with boring, plain charts. Data expressions that rely solely on visual channels will fan chong with blind or low-vision audience. Calculation on the Pan of People is about distilling the attributes of people and then diagnosing whether harmony is achieved.

## 5 Reinterpreting Design Guidelines in Visualization with Feng Shui.

In the Feng Shui compass, the compass itself is used to measure attributes, but there is also an "algorithm" or set of guidelines to determine whether the Feng Shui of something is good or bad. The human-centered design principles mentioned in the previous section—interpreted as fan chong—can be seen as one of these "algorithms" for our Feng Shui visualization compass. There are numerous design principles in the field of visualization, and many of them can serve as such algorithms for our compass—once they have been interpreted through the lens of Feng Shui theory.

Here, we use algebraic visualization design principles [12] as an example of translating visualization design guidelines into a Feng Shui narrative. We explore how these principles might be interpreted through the lens of Feng Shui. Note that Feng Shui is often considered a pseudoscience, and interpretations based on it are inherently speculative. As such, in this paper we will not attempt to justify the interpretations presented here—nor would it really be possible (so, essentially, just believe what this says, or risk bad luck).

The purpose of this interpretation is not to seek objective truth, but rather to construct a narrative that encourages belief in the usefulness of certain design choices.

Now let's walk through the Algebraic visualization design principles. Algebraic visualization design is a data-centered theory, so it describes how the Earth Pan (地盘) in the compass—and its relationships with other pans—should be computed. Representation Invariance states that it is the data itself, rather than spurious details of its representation, that should determine the impression made by a visualization. At its core, this principle is about determining the rotation state of the Earth Pan: if only spurious details change, the Earth Pan should not be rotated in a way that alters the overall impression of the visualization (Five Elements Pan). Unambiguous Data Depiction asserts that large changes in the data should be clearly visible. The underlying Feng Shui principle here is that changes in the Earth Pan (data) should be accurately reflected in the Five Elements Pan (general impression)—if the Earth Pan shifts, so too should the Five Elements; there should be no mismatch. Visual-Data Correspondence means that significant changes in the data should meaningfully correspond to noticeable changes in the visual impression, and vice versa. This further elaborates the linkage between the Earth Pan and the Five Elements Pan: a certain degree of rotation in the Earth Pan should correspond to a proportional rotation in the Five Elements Pan—and vice versa.

If we interpret some data visualization violations through the lens of the Feng Shui version of algebraic visualization design, we can map them directly to the three diagrams in the top row of Figure 1 in the original paper [12], each of which violates one of the core principles. The first diagram shows an abnormal disturbance of the Earth Pan. According to Feng Shui, "when the earth's energy is stable, life flourishes; when it is disturbed, decline follows." A disrupted Earth Pan symbolizes unstable earth energy and is considered highly inauspicious. The second diagram illustrates a case where changes in the Earth Pan are not reflected in the Five Elements Pan, leading to misaligned energy fields, imbalance between generating and overcoming forces, and disharmony of yin and yang—what Feng Shui calls a "broken situation of mismatch between name and reality (名实不符)." The third diagram shows the Earth Pan and the Five Elements Pan failing to remain in harmony as they change, which is also considered inauspicious.

## 6 Discussion

***Auspicious Artifacts to Improve Feng Shui of Visualization.*** Another important concept in Feng Shui, apart from the compass, is the use of auspicious artifacts (ji artifacts, 吉祥物)—objects strategically placed to attract good fortune, redirect energy, or dispel negative influences. This idea can also be applied to data visualization, where certain visual elements function in a similar capacity. A clear example of such a ji artifact in visualization is the use of annotations [20]. When placed appropriately, annotations can modulate the overall qi of the visualization, clarify complex areas, and serve as protective markers that guard against misinterpretation or cognitive overload—effectively "blocking sha qi" (挡煞气), or harmful informational energy. Much like a Feng Shui mirror might reflect away bad luck in a physical space, annotations can restore informational harmony within the visual field.

***Comfort Rituals to Alleviate Design Anxiety.*** Tackling a complex visualization can feel like wrestling chaotic *qi*, leaving designers anxious and grasping for certainty. Here, Feng Shui rituals step in as a psychological safety blanket. The simple act of arranging one's workspace or charts "just so"—however mystical it may seem—can provide a soothing illusion of control. In fact, studies show that even arbitrary rituals or superstitions can calm nerves and boost confidence, acting as a palliative against anxiety [7]. By believing their design follows auspicious principles (and thus *must* avoid misfortune), a designer might banish the specter of self-doubt and approach the task with a steadier hand. In psychological terms, this resembles a placebo effect: if a person trusts that their pre-design Feng Shui routine will bring good outcomes, that belief alone can melt away stress and sharpen performance. After all, when one feels the cosmic forces are on their side, the once-daunting data dashboard might suddenly seem less of a nightmare and more of an opportunity. In those moments, channeling a bit of *good-luck qi* can transform jittery hesitation into productive focus, making the creative process just a little more zen. Within the scope of this paper, we are unable to prove that this comforting power truly exists. That said, we would like to see it as a promising direction. Perhaps our current design cannot fully achieve this goal, and we may need to draw on more Feng Shui theories to make our compass more convincing, plus that the effect may differ with different cultural background.

***Symbolic Shortcuts and the Persuasive Power of Feng Shui.*** Feng Shui's real magic may lie in its knack for turning dry design rules into memorable stories. As a cognitive shortcut, a Feng Shui framing taps into our intuitive *System 1* thinking (thinking mode proposed by Daniel Kahneman [10])—the fast, automatic mode that runs on heuristics and gut feelings—rather than demanding slow, analytical reasoning from *System 2*. Psychologists note that our brains are "lazy" in this regard: System 2 will gladly default to whatever easy narrative System 1 serves up. A set of visualization guidelines wrapped in Feng Shui symbolism (yin-yang balance, five elements harmony, auspicious color choices, etc.) can thus be processed more fluently than a purely rational checklist. In essence, a Feng Shui narrative provides cultural resonance—it resonates with familiar beliefs and values—which means the guidelines "just feel right" on an emotional level. This folk-theoretic sheen can be surprisingly convincing. A designer might more readily accept a rule like "avoid chart junk to prevent bad feng shui" than an abstract decree about maximizing data-ink ratio, simply because the former comes with a built-in story and a hint of mystical consequence. By speaking to the imagination and aligning with cultural intuitions, the Feng Shui framework sneaks sound design principles into one's mind under the guise of legend. It's a pedagogical trojan horse: persuasive not by rigorous proof, but by engaging the audience's innate preference for narrative, symbolism, and a dash of ancient wisdom. In the grand duel of System 1 vs. System 2, if the feng shui-ified design rule wins, it does not win by force, but by charisma—making it far easier (and more fun) for people to nod along and incorporate good practices, no Guiana Chestnut required.

## Acknowledgments

This paper owns its existence to Yunyi Zhu, for helping draw the Feng Shui Compass in Figure 2, to Tingying He for dicussion on fun architectural Feng Shui examples. To ancestors and divine(s), for their wisdom and sacred guardianship.

# Appendix: Page-Sized Luo Pan Illustration

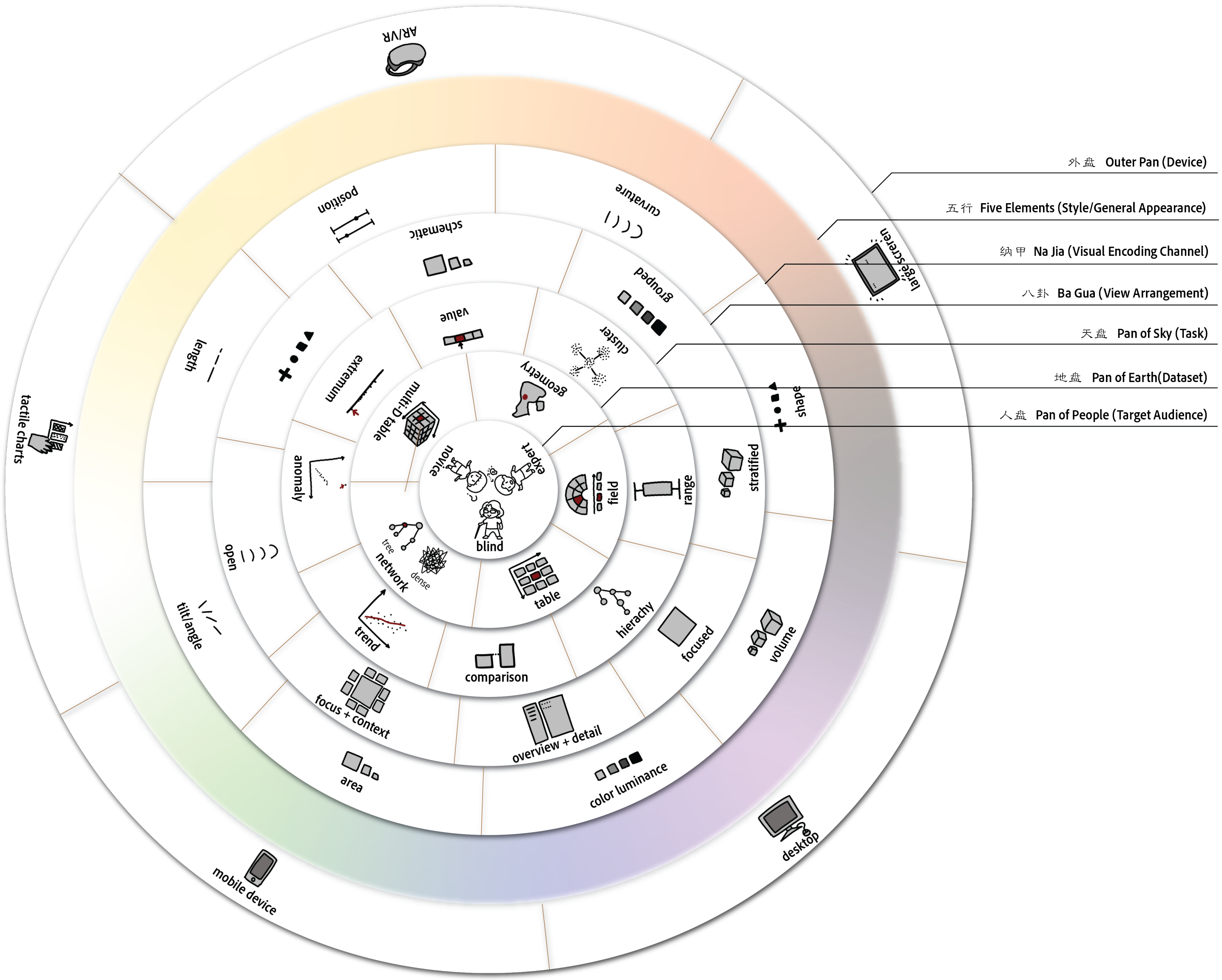